\documentclass[pdftex,twocolumn,epjc3]{svjour3}       

\RequirePackage[T1]{fontenc}

\RequirePackage{graphicx}
\RequirePackage{mathptmx}      
\RequirePackage{flushend}
\RequirePackage[numbers,sort&compress]{natbib}
\RequirePackage[colorlinks,citecolor=blue,urlcolor=blue,linkcolor=blue]{hyperref}

\usepackage[modulo]{lineno}
\journalname{Eur. Phys. J. C}

\begin{document} 


\title{Software compensation in particle flow reconstruction}

\author{Huong Lan Tran\thanksref{e1,addr1}\and
 Katja Kr\"uger\thanksref{addr1}\and
 Felix Sefkow\thanksref{addr1}\and
 Steven Green\thanksref{addr2}\and
 John Marshall\thanksref{addr2}\and
 Mark Thomson\thanksref{addr2}\and
 Frank Simon\thanksref{addr3}
}

\thankstext{e1}{e-mail: huong.lan.tran@desy.de}

\institute{Deutsches Elektronen-Synchrotron DESY, Notkestr. 85, 22607 Hamburg, Germany \label{addr1} \and
Cavendish Laboratory, JJ Thomson Avenue, Cambridge CB3 0HE, United Kingdom \label{addr2} \and
Max-Planck-Institut f\"ur Physik, F\"ohringer Ring 6, 80805 M\"unchen, Germany \label{addr3}}

\date{October 23, 2017}

\maketitle

\begin{abstract}
The particle flow approach to calorimetry benefits from highly granular calorimeters and sophisticated software algorithms in order to reconstruct and identify individual particles in complex event topologies. The high spatial granularity, together with analogue energy information, can be further exploited in software compensation. In this approach, the local energy density is used to discriminate electromagnetic and purely hadronic sub-showers within hadron showers in the detector to improve the energy resolution for single particles by correcting for the intrinsic non-compen\-sa\-tion of the calorimeter system. This improvement in the single particle energy resolution also results in a better overall jet energy resolution by improving the energy measurement of identified neutral hadrons and improvements in the pattern recognition stage by a more accurate matching of calorimeter energies to tracker measurements. This paper describes the software compensation technique and its implementation in particle flow reconstruction with the Pandora Particle Flow Algorithm (PandoraPFA). The impact of software compensation on the choice of optimal transverse granularity for the analogue hadronic calorimeter option of the International Large Detector (ILD) concept is also discussed. 
\end{abstract}

\section{Introduction}
\label{section:introduction}

Particle flow event reconstruction \cite{Brient:2002gh,pfaMorgunov,Thomson:2009rp} is one of the main drivers for the design of detector concepts at future energy-frontier  $e^{+}e^{-}$ colliders. In combination with highly granular calorimeter systems, this technique enables the precise reconstruction of multi-jet final states, which are of high relevance to TeV-scale physics. A particular goal of the detector designs for future linear colliders is the capability to separate W and Z bosons in hadronic decays. A separation of 2.3 - 2.6 $\sigma$ is achievable with a jet energy resolution of \mbox{3-4 \%} over a wide range of jet energies \cite{Thomson:2009rp}. Such resolutions cannot be achieved with the traditional calorimetric approach where all the jet components are exclusively measured in the electromagnetic and hadronic calorimeters (ECAL/HCAL). 

Particle flow algorithms make optimal use of all measurements of particles within a jet in the different subsystems of the experiment. For charged hadrons, which on average carry 60\% of the total energy, the most precise measurement is typically obtained in the tracker. Photons, accounting on average for 30\% of the total jet energy, and neutral hadrons with an average of around 10\% are only measured in the calorimeters. The optimal combination of all energy measurements has thus the potential to significantly reduce the impact of the intrinsically limited hadronic energy resolution of calorimeter systems in collider detectors.

\subsection{Detector systems designed for particle flow}
\label{subsection:IntroDetectors}

A key ingredient for the success of particle flow reconstruction is the capability to correctly match reconstructed tracks with calorimeter clusters to avoid missing or double counting energy. Detector designs optimised for particle flow thus feature highly granular calorimeters, a tracker providing  good momentum resolution with low material budget and a high magnetic field. The tracker and the calorimeters are located inside the magnet coil to avoid extended uninstrumented regions which would deteriorate the capability to associate different energy depositions of the same particle. 

All detector concepts proposed for possible future linear colliders such as ILC or CLIC are based
on the particle flow approach. They differ somewhat in overall size and magnetic field, and in the 
choice of the tracking system technology, but the proposed calorimeter systems are very similar. 
One example of a detector concept which fulfils this requirement is the International Large Detector (ILD) \cite{Behnke:2013lya} for the International Linear Collider \cite{Behnke:2013xla}. It uses a hybrid tracking system consisting of a large-volume gaseous Time Projection Chamber (TPC) tracking detector surrounded by silicon tracking layers and inner silicon strip and pixel detectors which together provide an excellent momentum resolution.

The electromagnetic and hadronic calorimeters in ILD as considered here have high longitudinal and transverse segmentation and are based on the silicon-tungsten ECAL \cite{Anduze:2008hq} and on the scintillator-steel analogue HCAL \cite{collaboration:2010hb} developed in the CALICE collaboration. The ECAL is composed of 30 layers with tungsten absorber and $5\times 5\ \mbox{mm}^{2}$ silicon cells. The tungsten absorber is ideal for the ECAL: it has a short radiation length and a small Moli\`ere radius which leads to compact electromagnetic showers, while featuring a large ratio of $\lambda_{I}$/$X_0$, which allows the ECAL to be almost ``transparent'' to hadrons. The first 20 ECAL layers have 2.1 mm thick tungsten absorber plates while the last 10 layers have 4.2 mm thick tungsten absorber plates. In total, the ECAL corresponds to 23 radiation lengths ($X_{0}$) and 0.8 nuclear interaction lengths ($\lambda_{I}$). For the HCAL the absorber material is stainless steel with a thickness of 20 mm per layer. The active elements consist of 3 mm thick $3\times 3\ \mbox{cm}^{2}$ plastic scintillator tiles directly read out with embedded Silicon Photomultipliers (SiPMs). The HCAL has a 48 layers, with a total thickness of  6$\lambda_{I}$. In addition to the technology choices outlined above, also scintillator/SiPM active elements and semi-digital RPC-based detectors are considered for the ECAL and the HCAL, respectively. These options are not included in the present study.

\subsection{Detector optimisation with PandoraPFA}
\label{subsection:intro_PFA}

Together with the detector design, the reconstruction software also plays a central role. The Pandora Particle Flow Algorithm (PandoraPFA) is the most widely used algorithm for detector benchmark and performance studies for future linear colliders. PandoraPFA was first introduced in 2009 \cite{Thomson:2009rp} and since then has been continuously maintained and improved. The tool provides a sophisticated pattern recognition for individual particle identification while remaining independent of the detector implementation \cite{Marshall:2015rfa}. Individual particles are reconstructed as so-called Particle Flow Objects (PFOs), combining the information obtained from different sub-detectors. The two-shower separation in the calorimeters, a key performance driver of the algorithm, has been studied both in simulations and on CALICE data \cite{Adloff:2011ha}, finding good agreement with little dependence on the precise shower physics model and giving confidence in the simulation studies used to develop and evaluate the algorithm in a collider detector setting. 

\begin{figure}[!htbp]	
  \begin{center}	
    \includegraphics[width=0.42\textwidth]{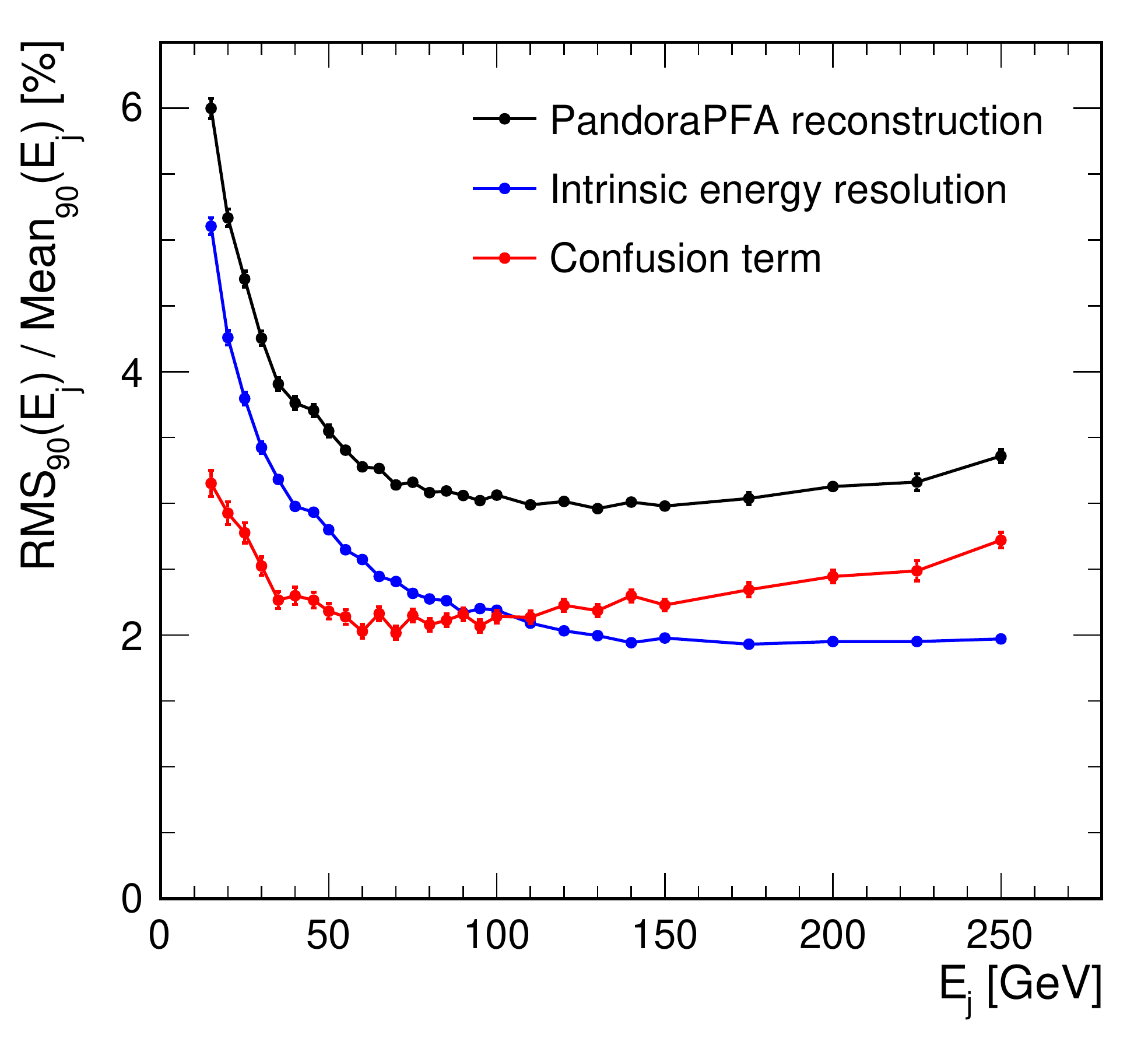}
  \end{center}	
  \caption{Jet energy resolution, given by $\textrm{RMS}_{90}/\textrm{Mean}_{90}$ (see Section \ref{section:performance_results} for details of the jet energy resolution definition), obtained with the PandoraPFA algorithm for the ILD detector concept. The intrinsic energy resolution is obtained by using the information of which particle is responsible for a given energy deposition in the calorimeter from Geant4 in the pattern recognition, ensuring correct matching of particle tracks and calorimeter clusters. The confusion term is the contribution to the resolution originating from mistakes in the assignment of energy depositions to reconstructed particles, defined as the quadrature difference between the total and the intrinsic energy resolutions.}
\label{fig:JER_vs_Ej}
\end{figure}

PandoraPFA has been used to optimise the ILD detector design. For the optimisation of the calorimeters, the jet energy resolution is taken as the main parameter.  Figure ~\ref{fig:JER_vs_Ej} shows the jet energy resolution obtained for di-jet events with jets originating from light quarks ({\it{u, d, s}}) in the barrel region ($\vert \mathrm{cos}\theta \vert\, < \, 0.7$) of the ILD detector. In this analysis, cell energy truncation is used in the reconstruction as discussed in more detail in Section \ref{subsubsection:cell_energy_truncation}. For the 50 to 250 GeV energy range, the jet energy resolution is below 4\%. 

The contributions of different effects to the total jet energy resolution are also studied to identify the main performance drivers. The intrinsic particle energy resolution is determined using \textit{perfect pattern recognition} algorithms. These algorithms use the information of the location of energy depositions belonging to a given particle available in Geant4 to provide a correct assignment of calorimeter energy deposits to PFOs. In this procedure, the matching of calorimeter cells with energy depositions to clusters and tracks is "cheated". Each calorimeter cell is assigned to the MC particle which is responsible for the largest fraction of the energy in that cell. Given the high granularity of the calorimeters, the possible ambiguity in this assignment has negligible effects on the overall energy assignment. The detector response itself is not altered with respect to the regular reconstruction, taking into account realistic detector effects as discussed in section \ref{subsection:sim_reco_framework}. This effectively removes the influence of imperfect pattern recognition and energy assignment, thus leaving only the intrinsic energy resolution of the detector as a source of uncertainty. 

The confusion term, which is defined as the quadrature difference between the total and intrinsic energy resolution, represents the contribution of mistakes made in the reconstruction of the PFOs. The confusion term receives contributions both from calorimeter energy lost in the reconstruction and from double-counting of energy. A typical example for the former is the case where calorimeter hits belonging to a photon are not properly resolved from a cluster belonging to a track and therefore the photon energy is not accounted for. Double-counting of energy can occur if a part of a charged hadron shower is mis-identified as a separate cluster. In this case this cluster's energy is counted twice in the jet energy reconstruction. For jet energies below 100 GeV, the intrinsic calorimetric energy resolution is the main driver, whereas for higher energy jets starting from 100 GeV the confusion term starts to be dominant.

\subsection{The software compensation concept}
\label{subsection:intro_softwarecompensation}

Even though the particle flow reconstruction approach reduces the influence of the calorimeter energy resolution on the overall jet reconstruction performance, resolution still plays an important role. It directly influences the contribution of the neutral hadron energy measurement on the jet energy, and is also relevant in the matching of calorimeter clusters to reconstructed tracks. Optimising the hadronic energy resolution of the calorimeter system is thus also highly relevant in detectors designed for particle flow.

In most experiments, the calorimeters are non-com\-pen\-sating, which means that the detector response for electromagnetic showers is typically higher than for hadronic showers ($e$/h $>$ 1). The non-compensation arises from the complexity of the hadronic showers, which consist of an electromagnetic and a hadronic component. The electromagnetic component originates primarily from the production of $\pi^{0}$ and $\eta$ particles in hadronic showers, which instantly decay into photon pairs, resulting in electromagnetic cascades in the detector. The hadronic component contains unaccessible processes, such as nuclear target recoil in which the generated energy usually does not reach the active medium, binding energy losses from nuclear breakup or late neutron capture which happens with substantial delay, beyond the integration time of the detector electronics. These undetectable processes lead to a response that fluctuates strongly from event to event and is on average lower than that of electromagnetic showers, where in principle the full energy is detectable. Since the electromagnetic fraction strongly fluctuates from event to event, the difference in response for the two components results in a degradation of the energy resolution. In addition, the average electromagnetic fraction increases with the energy of the incoming particle, which results in a non-linear calorimeter response \cite{Gabriel:1993ai}. 

Compensating calorimeters ($e$/h $\approx 1$) can thus improve the energy reconstruction and the resolution for hadrons. Compensation can be reached with specific detector designs using heavy metal absorbers and hydrogenous active elements in a specific ratio \cite{Wigmans:1986pe}. These requirements however impose strong constraints on the geometry and uniformity of the calorimeter system. On the other hand, a compensation of the detector response is also achievable for intrinsically non-compensating calorimeters using signal weighting techniques, here referred to as software compensation. These techniques exploit the fact that the electromagnetic components tend to have a higher energy density compared to purely hadronic components, since their evolution is characterised by the radiation length, which is substantially shorter than the nuclear interaction lengths in typical calorimeter materials. The local signal amplitude in segmented calorimeters can thus be used to improve the energy resolution through a reduction of shower-to-shower fluctuations in visible energy based on appropriate weighting of the signals of individual readout channels. This was first demonstrated in the iron-scintillator calorimeter of CDHS \cite{Abramowicz:1980iv}, was used in collider experiments such as H1 \cite{Andrieu:1993tz} and ATLAS \cite{Cojocaru:2004jk}, and successfully applied to the highly granular CALICE analogue hadron calorimeter \cite{Adloff:2012gv}. 

In this paper, we discuss the technical aspects of the implementation of software compensation in PandoraPFA, following the so-called \textit{local software compensation} technique used in CALICE, followed by a study of the performance of this technique for jet energy reconstruction in ILD detector concept, and an investigation of the impact of the transverse granularity of the hadron calorimeter on the jet energy resolution.

\section{Implementation of software compensation in the particle flow reconstruction}
\label{section:Implementation}

The technical framework used to study software compensation in the ILD detector concept can be sub-divided into three discrete areas, connected to different aspects of the problem, as discussed below. The first is the simulation and reconstruction of the events used in the study, the second is the implementation and the calibration of the software compensation on the calorimeter cell level, and the third one is the integration of the technique into the PandoraPFA particle flow algorithm. 

\subsection{Simulation and reconstruction framework}
\label{subsection:sim_reco_framework}

The present study is performed using Monte Carlo samples of single hadrons for calibration and jets to test the performance of the algorithm. The simulations use the default settings of ILC physics simulations. Jet fragmentation and hadronisation are based on the PYTHIA \cite{Sjostrand:2000wi} generator tuned to the fragmentation data from the OPAL experiment \cite{Alexander:1995bk}. The generated physics events were simulated in the Mokka framework \cite{MoradeFreitas:2002kj} using Geant4 \cite{Agostinelli:2002hh} version 9.6 and a realistic model of the ILD detector concept outlined in section~\ref{subsection:IntroDetectors} (ILD\_o1\_v06). For the simulation of particle showers in the detector material, the QGSP\_BERT \cite{Dotti:2011zz} physics list of Geant4 was used. The detector simulation includes a number of instrumentation effects such as the dead regions around the silicon detector pads in the ECAL, gaps between modules in the ECAL and HCAL and an implementation of Birk's law \cite{Birks:1964zz} to model saturation effects in the scintillator response in the HCAL as well as smearing of the HCAL cell energies to account for photon statistics and detector noise, amplitude cuts for noise rejection in the reconstruction and timing cuts to realistically model the finite integration time of the read-out electronics. For the reconstruction of the events, the full digitisation and reconstruction chain in the Marlin framework \cite{Gaede:2006pj} of ILCSoft version v01-17-07, combined with PandoraPFA version v02-09-00 is used. The simulation and reconstruction correspond to the standard ILD software and include detailed modelling of detector effects, validated with test beam data. Only some effects corresponding to specific design choices of ILD have been removed, e.g. some services or dead areas due to guard rings, to preserve the generic nature of the results. It was verified that these effects do not alter the resolutions by more than 5\% relative.

\subsection{Implementation and calibration of software compensation weights}
\label{subsection:SC_weight_definition}

The present study is based on the local software compensation technique developed for the highly granular calorimeters of CALICE, as described in \cite{Adloff:2012gv}. This technique uses cell-by-cell weights in the calorimeter, with the weights a function of the local energy density $\rho$. Here, $\rho$ is defined as the hit energy in GeV divided by the cell volume in units of $\mbox{1000\, cm}^{3}$, taking the volume of the scintillator and the absorber and other passive materials into account. In this definition, a standard HCAL calorimeter cell has a volume of 23.85 cm$^3$. The volume normalisation in principle allows for treating cells with different sizes. In 
the study here, the HCAL has uniform granularity, so this does not affect the performance. 

\begin{figure}[!htbp]	
  	\begin{center}	
    	\includegraphics[width=0.42\textwidth]{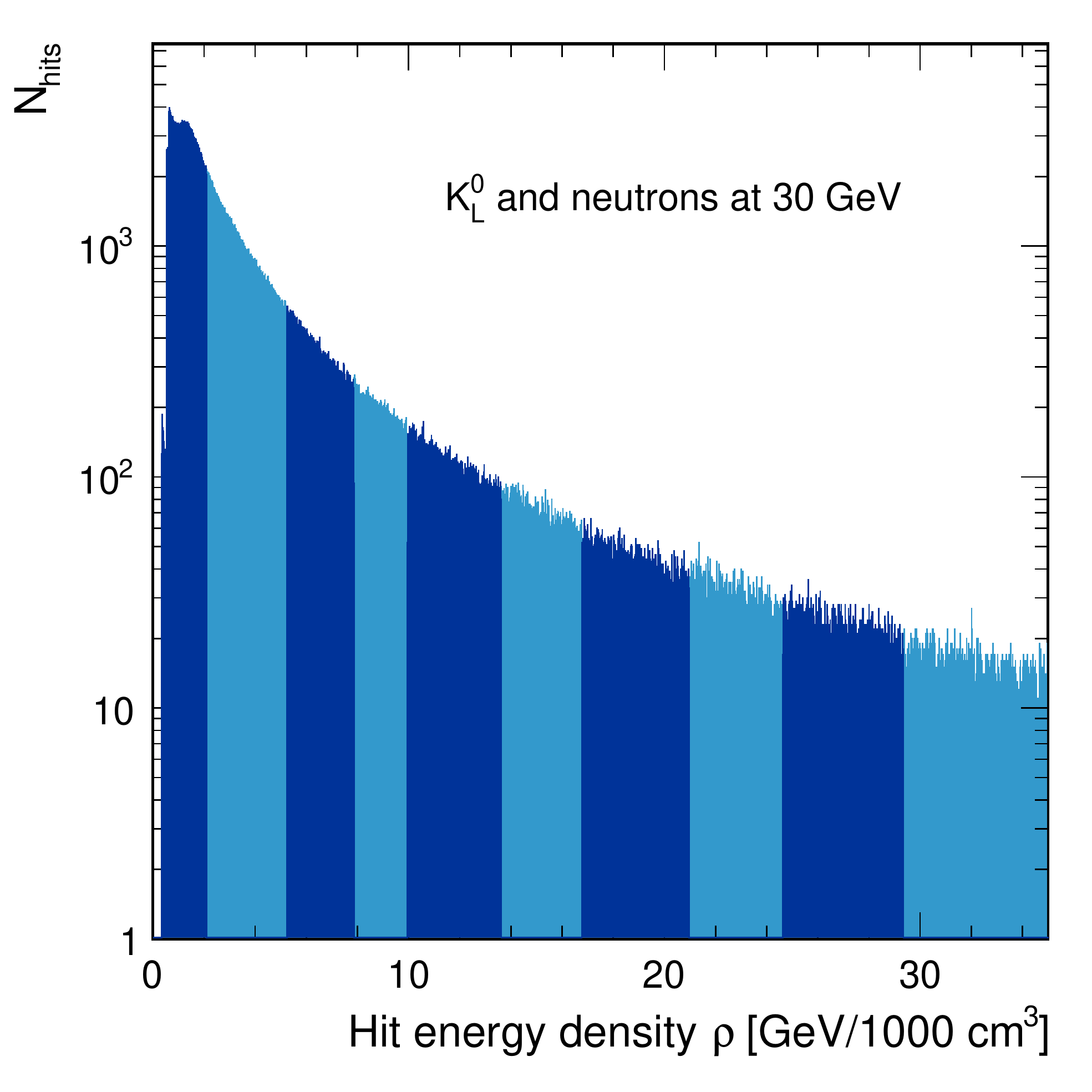}
  	\end{center}	
  	\caption{Distribution of the hit energy density for a sample of $K^{0}_{L}$ and neutrons with 30 GeV energy. One MIP per cell corresponds to approximately 1 GeV / 1000\,cm$^{3}$. The differently shaded areas show the subdivision into energy density bins used for the software compensation.}
  	\label{fig:HitEnergy}
	\end{figure}

A typical distribution of the hit energy density is shown in Fig.~\ref{fig:HitEnergy}. On this scale, a minimum-ionising particle corresponds to approximately 1 GeV / $\mbox{1000\, cm}^{3}$, with the first bin extending to 2 MIPs per cell. If a hit has higher energy density, it is more likely to originate from to the electromagnetic component of the shower. Conversely, a hit with lower energy density is more likely to be of purely hadronic origin. In order to reach compensation (ratio $e$/h close to 1), the response to different hits with different energy densities is re-weighted, with hits with a higher density receiving a smaller weight than those with lower energy density.

The energy of calorimeter clusters are  computed as:
\begin{eqnarray}
\label{eq:SC_energy_computation}
&& \mathrm{E}_{\mathrm{SC}} = \sum_{\mathrm{hits}} \mathrm{E}_{\mathrm{ECAL}} + \sum_{\mathrm{bin}\ i} (\mathrm{E}^{i}_{\mathrm{HCAL}}\times \omega(\rho_{i})) \\
&& \mbox{with}\ \mathrm{E}^{i}_{\mathrm{HCAL}} = \sum_{\mathrm{hits}\,\in\, \mathrm{bin}\ i} \mathrm{E}_{\mathrm{hit}}, 
\end{eqnarray}
where $\omega(\rho)$ gives the weight of a hit as a function of its energy density $\rho$. 	

This weight function are expressed by the sum of an exponential function and a constant to enforce a monotonically falling behaviour with energy density \cite{Adloff:2012gv},
\begin{equation}
\label{eq:weight_exp_function}
\omega(\rho) = p_{1} \, \mathrm{exp}(p_{2}\, \rho) + p_{3}, 
\end{equation}
where the parameters $p_1$, $p_2$ and $p_3$ depend on the energy of the particle to take overall changes of the shower structure and density profiles with particle energy into account. Here, $p_2$ is negative while the other two parameters are positive, resulting in lower weights for higher hit energies. To extract this weight function, the energy density is divided into 10 bins as illustrated by the differently shaded areas in Fig.~\ref{fig:HitEnergy}. The sum of hit energies of each bin is then weighted by $\omega(\rho)$, with $\rho$ being the centre of the bin, meaning that each hit within a given density bin is weighted with the same weight. 

As discussed in \cite{Adloff:2012gv}, the choice of the number of bins is a compromise between high granularity of the weights and stability of the algorithm. In the CALICE study, the bin border were chosen such that each bin contributes approximately the same to the total energy of typical events. Since the present study is performed for a wide variety of cell sizes, the bin borders were re-optimized to obtain stable convergence of the training procedure for all cell sizes considered. In general, the performance of the algorithm does not depend strongly on the exact choice of the bin borders or on the number of bins. 

By design, the exponential parametrisation of the dependence of the weight on local energy density provides a good description of the expected weight behaviour: it gives larger weights for low density hits, which are likely to be of hadronic origin, and smaller weights for high density hits, which are more likely generated in the electromagnetic component. Since the electromagnetic fraction and the overall particle density in a hadronic shower changes with the energy of the incoming hadron, it is expected that the weights change with particle energy as well. The parameters $p_{1}$, $p_{2}$ and $p_{3}$ therefore have to take this dependence into account. Following \cite{Adloff:2012gv}, the parameters are expressed as functions of the unweighted energy sum $\mathrm{E}_{sum}$ of the cluster. The functional forms used to model the energy dependence is determined in a phenomenological approach, based on the observed energy dependence of the weight parameters,
\begin{eqnarray}
p_{1} &=& p_{10} + p_{11} \times \mathrm{E}_{sum} + p_{12} \times \mathrm{E}^{2}_{sum} \label{eq:weightparamater1}\\ 
p_{2} &=& p_{20} + p_{21} \times \mathrm{E}_{sum} + p_{22} \times \mathrm{E}^{2}_{sum}  \label{eq:weightparamater2}\\
p_{3} &=& \frac{p_{30}}{p_{31} + e^{p_{32}\times \mathrm{E}_{sum}}}.  \label{eq:weightparamater3},
\end{eqnarray}
where $p_{i0}$, $p_{i1}$ and $p_{i2}$ are numerical constants used to model parameter $p_{i}$. 

In the present study, the software compensation is applied to all particle showers in the main calorimeter system of ILD (barrel and endcap), however, only energy deposits in the HCAL are re-weighted. Additional work needs to be done to include the ECAL, but since the energy deposition from hadronic showers is primarily in the HCAL, the application of software compensation within this region is sufficient to realise the majority of the potential performance gain, while at the same time keeping the complexity of the calibration and application of the technique rather low. Energy deposits outside the HCAL do contribute to the raw cluster energy that determines the weights to be applied, and are counted in the final energy estimate with unit weight.

The software compensation weights are determined by the minimisation of a  $\chi^{2}$-like function
\begin{equation}
\label{eq:chi2_SC_weights}
\chi^{2} = \sum_{\mathrm{events}} \frac{(\mathrm{E}_{\mathrm{hadron}} - \mathrm{E}_{\mathrm{SC}})^{2}}{(0.5)^2 \, \mathrm{E}_{\mathrm{hadron}} \mathrm{GeV}}
\end{equation}
in a dedicated calibration procedure. The energy dependence of the weights is obtained by carrying out the minimisation of this function using calibration events with single neutral hadrons simulated for a range of energies, as discussed below. In this process, equal statistics of each energy point are required to avoid biases towards particular energies.  An equal weight of the different simulated energies in the overall $\chi^2$ is ensured by normalising the deviation of the reconstructed from the true energy by the approximate single particle stochastic resolution term of $50\% \sqrt{\mathrm{E}_{\mathrm{hadron}}\, \mathrm{GeV}}$ for a CALICE - style analogue hadron calorimeter \cite{Adloff:2012gv}.

\begin{figure}[!htbp]	
  	\begin{center}	
    	\includegraphics[width=0.42\textwidth]{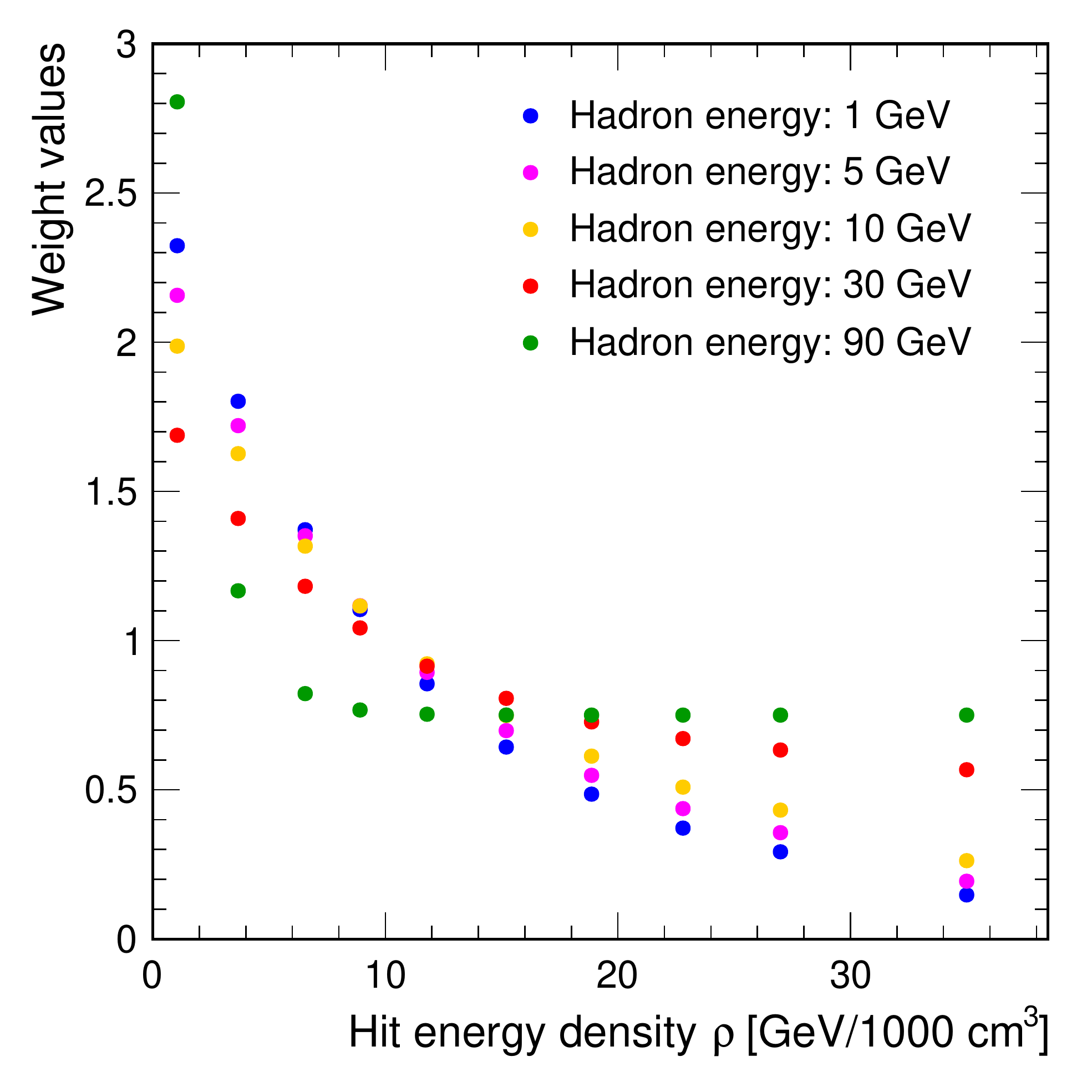}
  	\end{center}	
  	\caption{Software compensation weights as a function of hit energy density for different hadron energies, derived from Eq. \ref{eq:SC_energy_computation}.}
  	\label{fig:SC_weight_values}
	\end{figure}

For this calibration, a sample of single neutral hadron (\textit{neutron} and $K^{0}_{L}$) events with energies ranging from 10 GeV to 95 GeV is used. The hadrons originate from the interaction point and only events with one cluster fully contained in ECAL and HCAL are selected. This ensures that the weight calibration is not biased by events that have clusters not matching the expected original energy. The fraction of events rejected by this requirement is approximately 30\%, primarily due to events with small shower fragments not correctly associated to the main PFO. Fig.~\ref{fig:SC_weight_values} shows the derived weight values based on Eq. \ref{eq:weight_exp_function} as a function of the hit energy density for different hadron energies. Note that Eqs. \ref{eq:weightparamater1} -- \ref{eq:weightparamater3} are used here to extrapolate the energy dependence of the weights beyond the range of the calibration points to lower energy, resulting in sensible values also at 1 GeV. This is an important property for the overall stability of the algorithm when applying to jets with particles with a wide range of energies. 

The mean energy of neutral hadrons entering the calorimeter system ranges from 5.6 GeV to 16 GeV for light-quark jets with an energy from 45.5 GeV to 250 GeV, with a somewhat lower average energies for charged particles. At the same time, the energy distributions have a pronounced tail extending to high energies, up to the full jet energy. This underlines the importance of a robust performance at both low and high particle energies.

\subsection{Implementation in PandoraPFA}
\label{subsection:implementation}

PandoraPFA uses a multi-algorithm approach to calorimeter pattern recognition, with iterative steps building up the calorimeter clusters in combination with tracking information to arrive at the final PFOs. The energy of calorimeter clusters primarily enters at two stages in the process. The "reclustering" algorithm uses the matching of the calorimeter cluster energy estimation and the momentum of the associated track for charged particles to guide an iterative re-clustering. It splits or merges calorimeter clusters with energies inconsistent with an associated track to try to achieve a better energy-momentum consistency. In the final building of the PFOs the cluster energy is used for neutral particles, and taken into account in combination with the track momentum for charged particles. 

For the determination of the weights used by the software compensation algorithm, the uncorrected cluster energy is used. In principle, for the test of a given track cluster matching hypothesis, the measured momentum of the track could also be used in the choice of weights for the compensated energy estimate of the cluster. However, in a dense jet environment, with several tracks to be tested for the same cluster, this leads to additional complications and combinatorial overhead. Since studies on test beam data have shown that very little gain is to be expected for a scintillator-based analogue calorimeter when using the true rather than the measured cluster energy \cite{Adloff:2012gv}, this has not been implemented. 

The software compensation procedure outlined in the preceding subsection is implemented in the PandoraPFA particle flow algorithm making use of the features provided by the Pandora software development kit (SDK) \cite{Marshall:2015rfa}, which provides a flexible environment for developing and running pattern recognition algorithms as part of the particle flow reconstruction. In the present study, the software compensation is implemented as a user-defined "plugin" cluster energy estimator. For a set of input calorimeter hits, grouped into a cluster, the plug-in provides an estimate of the energy, making use of software compensation. The pattern recognition algorithms can access this information by requesting the corrected hadronic energy for a cluster. 

In the present study, we consider three different cases for the cluster energy reconstruction:  
\begin{enumerate}
\item The default energy estimation without the use of SC as a reference;
\item the application of the SC-based cluster energy correction in the final PFO construction stage only;
\item the full use of software compensation throughout the reconstruction, allowing the clustering algorithms within PandoraPFA to access improved energy estimators to inform pattern recognition decisions.
\end{enumerate}

\section{Software compensation performance}
\label{section:performance_results}

In this section, the effect of the software compensation technique on the energy reconstruction of single particles and hadronic jets is discussed. 

As in other PandoraPFA-related studies, the energy resolution of jets, used as a performance metric, is defined as $\mathrm{RMS}_{90}/\mathrm{Mean}_{90}$. $\mathrm{RMS}_{90}$ and $\mathrm{Mean}_{90}$ are the standard deviation and the mean of the reconstructed energy distribution over the smallest range of total reconstructed energy which contains 90\% of all events. For the single particle energy resolution, the $\mathrm{RMS}$ and $\mathrm{Mean}$ values are quoted from a gaussian fit on this 90\% integral range. This is done to make the single particle resolutions directly comparable to the energy resolution of the CALICE AHCAL \cite{Adloff:2012gv}, which is determined from gaussian fits to the reconstructed energy.

All plots in the following show the comparison of three different reconstruction schemes: the standard reconstruction without energy corrections based on SC, SC applied only to neutral hadrons after pattern recognition and with SC also used at the stage of pattern recognition in particle flow.

\subsection{Single particle energy resolution}
\label{subsection:single_particle_energy_resolution}

The software compensation weights are applied first to single particle samples of $K^{0}_{L}$ and neutrons from 10 to 95 GeV. The reconstructed energy is computed using the energy sum of all PFOs in the event. As discussed in section \ref{subsection:SC_weight_definition}, approximately 30\% of the single particle events are reconstructed with more than one PFO. In the majority of these cases, only small cluster fragments are not correctly associated to the leading PFO. To study the energy reconstruction free from issues of pattern recognition, all reconstructed PFOs in the event are considered here.  Decays of $K^{0}_{L}$ in the detector volume are not explicitly excluded, but the contribution of such processes is negligible. 

\begin{figure}[!htbp]	
  	\begin{center}	
    	\includegraphics[width=0.42\textwidth]{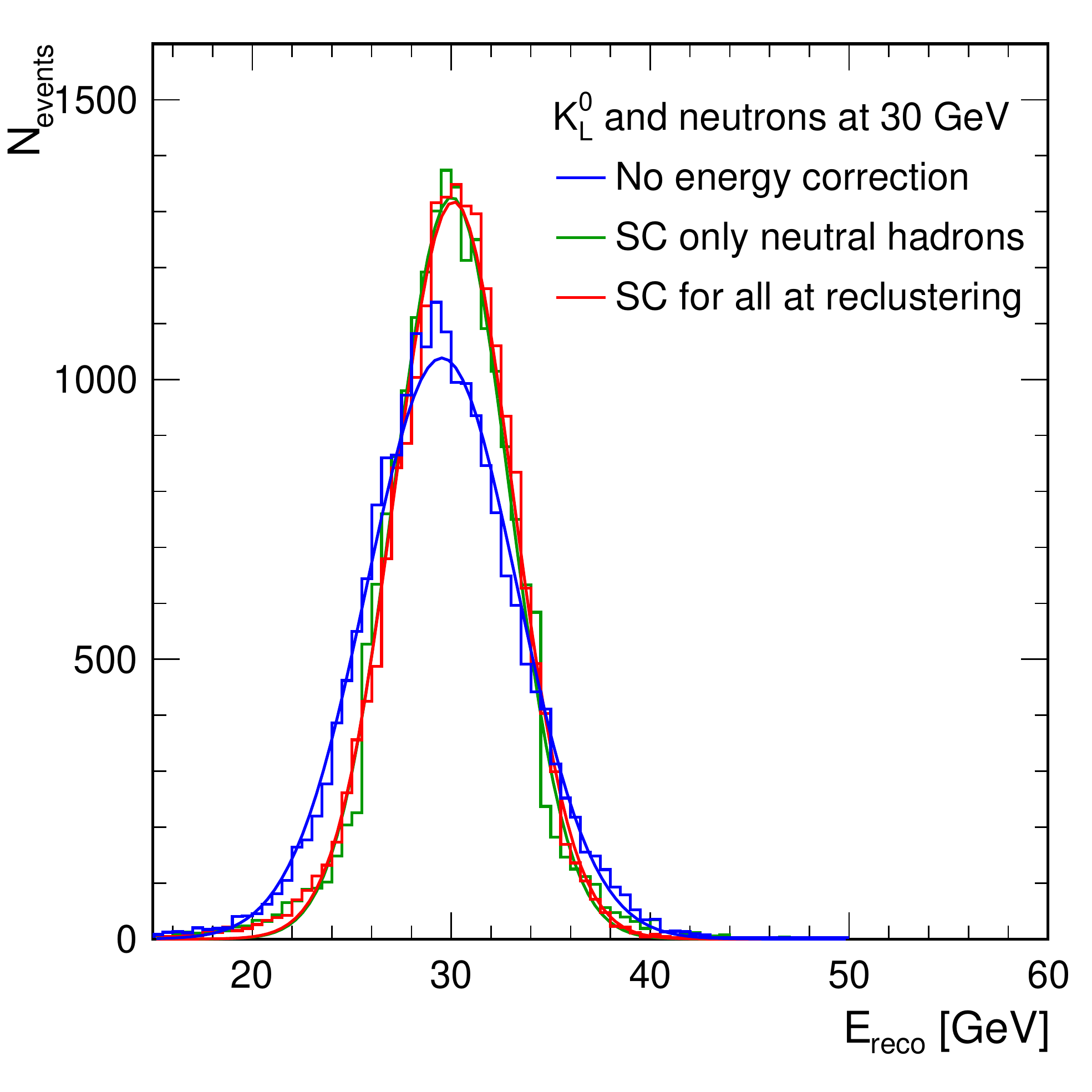}
  	\end{center}	
  	\caption{Distribution of energy sum of all PFOs for neutrons and $K^{0}_{L}$ at 30 GeV: without energy correction (blue), when applying SC for neutral cluster at PFO level (green) and when applying SC at the reclustering step for all clusters (red).}
  	\label{fig:/EnergyDistributionK0}
	\end{figure}

When applying SC, the distribution of the energy sum becomes more gaussian and narrows considerably, as shown in Fig.~\ref{fig:/EnergyDistributionK0}. The fluctuations in the reconstructed energy are considerably reduced since events with a high electromagnetic content and a correspondingly higher visible energy are down-weighted, while those dominated by purely hadronic activity are reconstructed with higher average weights. When applying SC additionally at the reclustering step no significant change in the improvement is seen. This is expected, since for neutral hadrons there is no reconstructed track pointing to the calorimeter cluster, which could be used to guide the pattern recognition. Subtle differences between the two methods arise due to threshold effects in the PFO creation, since the application of SC during the reclustering changes the hit energies before the final PFOs are created, while the application of SC to neutral hadrons only changes the energy estimates for neutral hadrons after the pattern recognition is completed. 

\begin{figure}[!htbp]	
  	\begin{center}	
    	\includegraphics[width=0.42\textwidth]{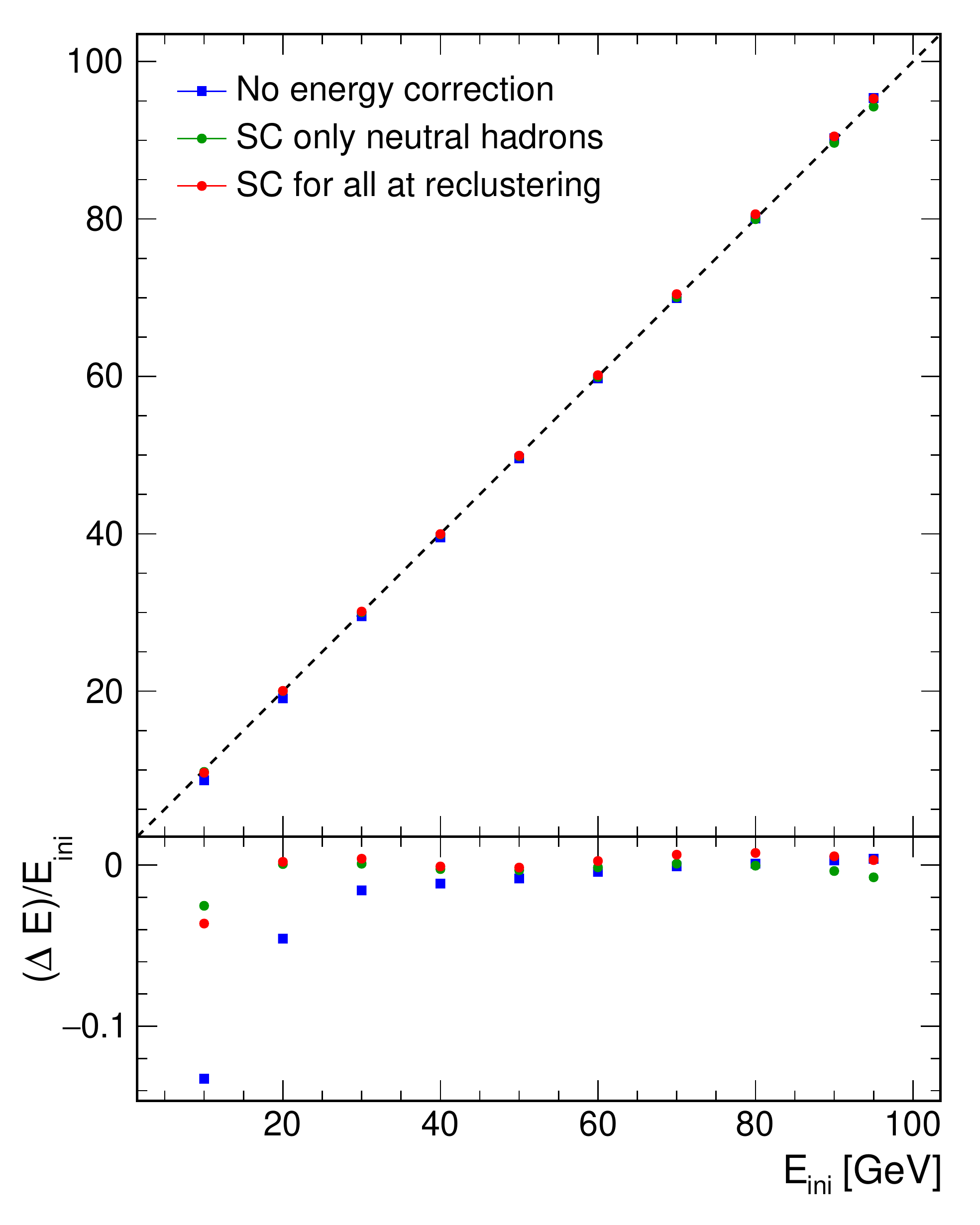}
  	\end{center}	
  	\caption{Mean reconstructed energy for single neutrons and $K^{0}_{L}$ using the sum of all PFOs: without energy correction (blue), when applying SC to neutral clusters at the PFO level (green) and when applying SC also during pattern recognition for all clusters (red). The dashed line shows the $\mathrm{E}_{reco} = \mathrm{E}_{ini}$ relationship.}
  	\label{fig:EnergyReconstructionK0}
	\end{figure}

Fig.~\ref{fig:EnergyReconstructionK0} shows the mean reconstructed energy of single particles as a function of energy. Software compensation improves the linearity of the energy response of the detector considerably, with deviations below 4\% over the full energy range considered. 

\begin{figure}[!htbp]	
  	\begin{center}	
    	\includegraphics[width=0.42\textwidth]{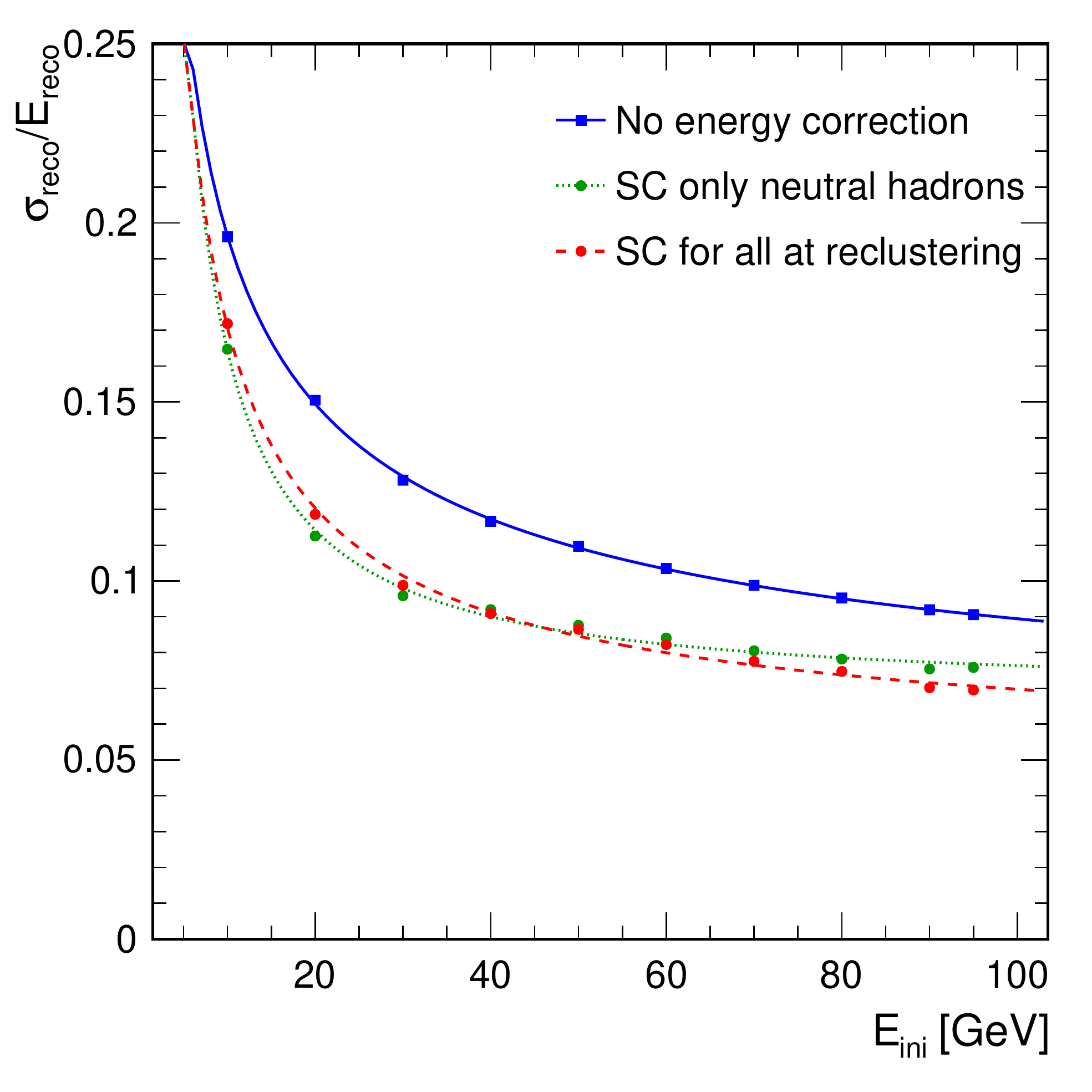}
  	\end{center}	
  	\caption{Single particle energy resolution without energy correction (blue), when applying SC for neutral cluster at PFO level (green) and when applying SC at the reclustering step for all clusters (red).}
  	\label{fig:EnergyResolutionK0}
	\end{figure}

The energy resolution of single particles is clearly improved over the whole range of energy from 10 to 95 GeV, as demonstrated in Fig.~\ref{fig:EnergyResolutionK0}. The relative improvement ranges from $\sim$13\% to $\sim$26\%. This result is compatible with the observations made by the CALICE collaboration when applying the software compensation technique to charged pion test beam data with an energy from 10 GeV to 80 GeV \cite{Adloff:2012gv}. 

\subsection{Jet energy resolution}
\label{subsection:jet_energy_resolution}

The performance of software compensation in jet energy reconstruction is studied in simulated di-jet events  $e^{+}e^{-} \rightarrow (Z/\gamma)^{*} \rightarrow \mbox{q}\bar{\mbox{q}}$ considering only decays  to light quarks (i.e. q = u,d,s). The two jets are produced back-to-back with equal energy. The total energy of the two-jets system $\mathrm{E}_{\mathrm{jj}}$ is reconstructed and the jet energy resolution is obtained by multiplying the relative total energy resolution by $\sqrt{2}$. Note that this procedure ignores effects of jet finding on the jet energy reconstruction and resolution. 

Four different centre-of-mass energies are considered:  \mbox{$\sqrt{s}$ = 91, 200, 360} and 500 GeV. A cut on the polar angle of the generated quarks ($|\mathrm{cos}\theta_{q}|, |\mathrm{cos}\theta_{\bar{q}}| <$ 0.7) is applied to select jets predominantly in the barrel region of the detector, de-emphasising the barrel-endcap transition region. 

In these jets, the typical number of long-lived particles reaching the calorimeters ranges from 22 for a jet energy of 45.5 GeV to 42 for an energy of 250 GeV. Approximately 50\% of these particles are charged particles, 45\% are photons and 5\% are neutral hadrons. The neutral hadrons carry 13\% of the total jet energy on average for all jet energies considered here. 

	\begin{figure*}[!htbp]
  	\centering
  	\includegraphics[width=0.42\textwidth]{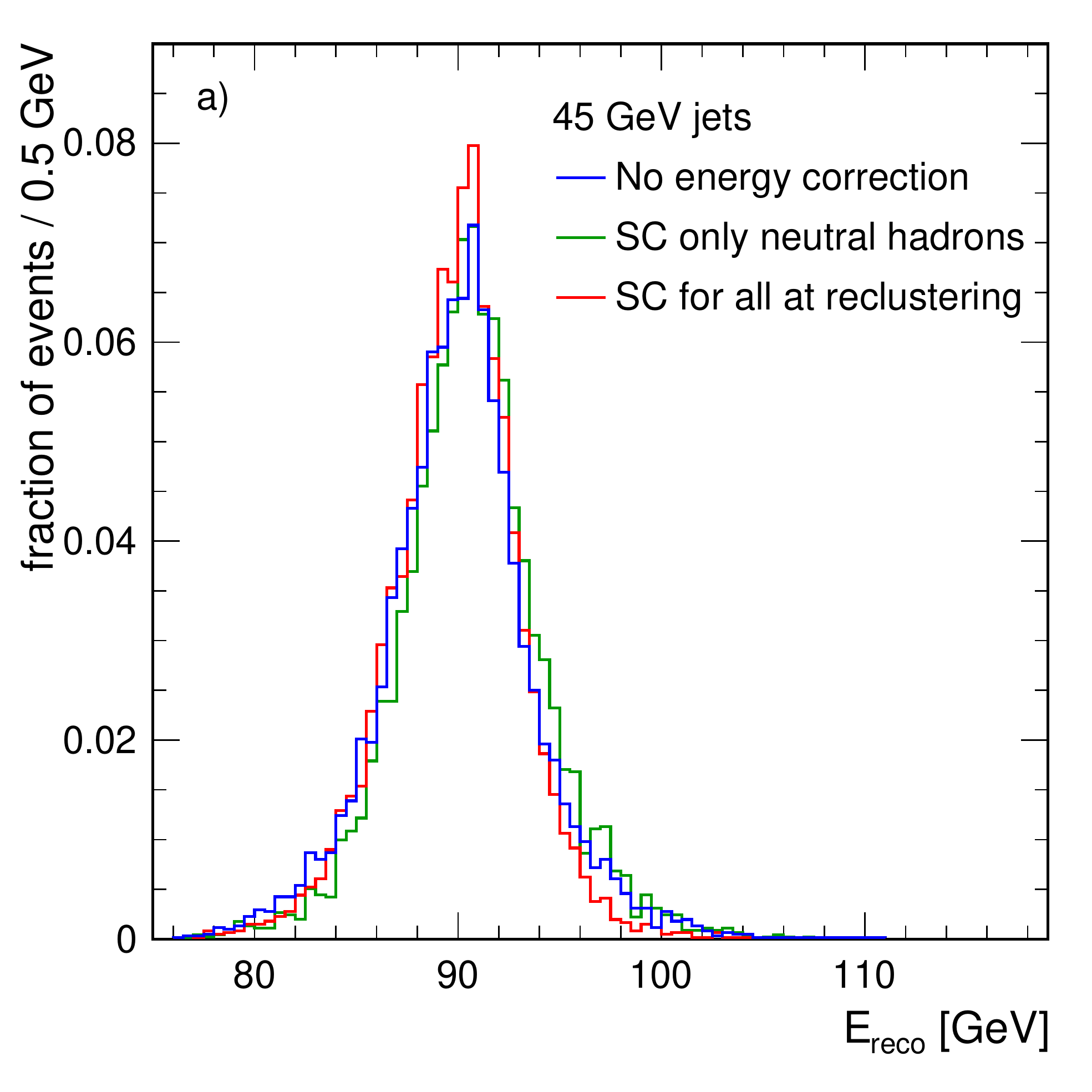}
	  \hfill
 	\includegraphics[width=0.42\textwidth]{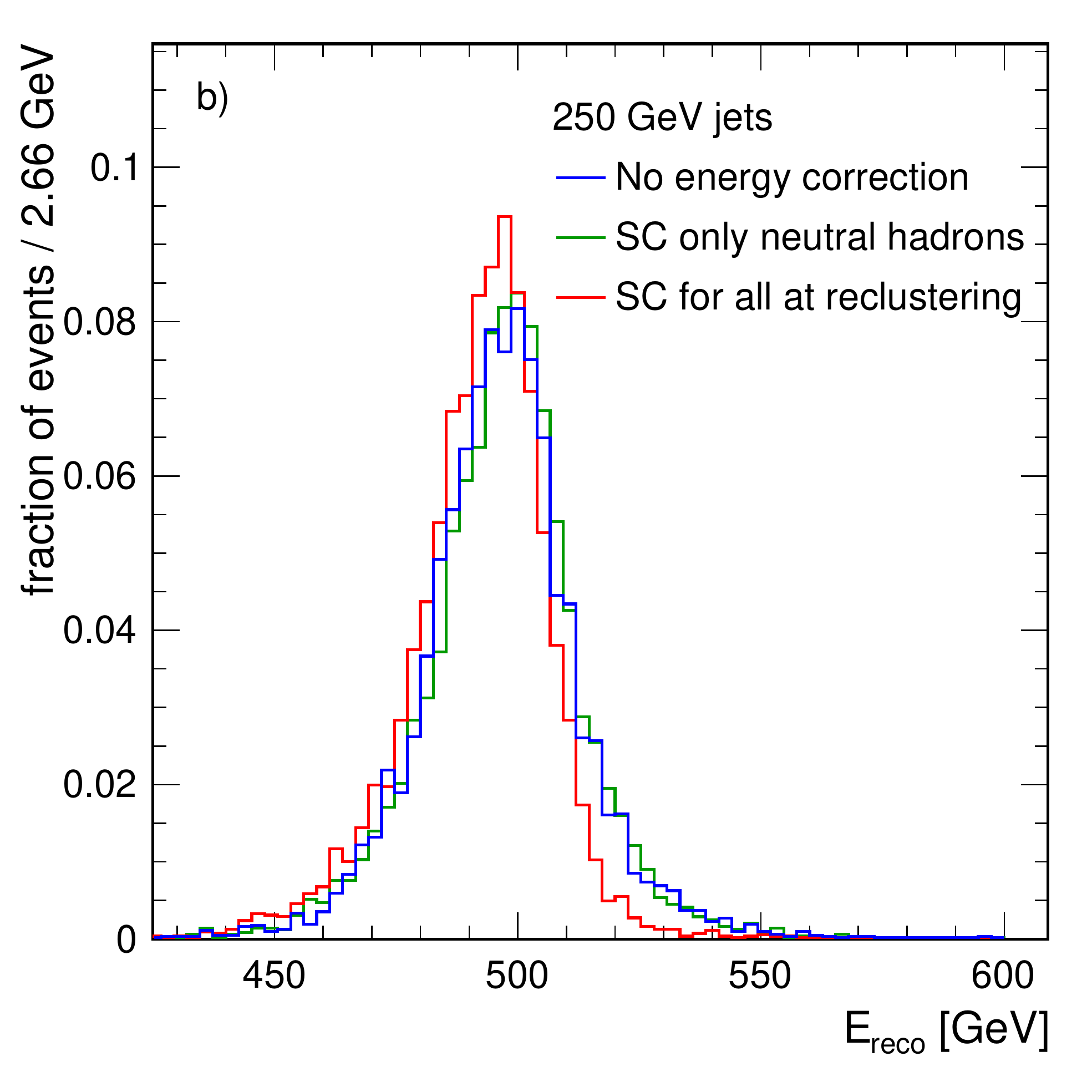}
	  \caption{Normalised distribution of the energy sum per event for two different jet energies: a) 45 GeV and b) 250 GeV, for all three scenarios of energy correction studied here.}
	  \label{fig:E_PFO}
	\end{figure*}

Fig.~\ref{fig:E_PFO}  shows the normalised distributions of the reconstructed energy sum per event for two different jet energies, 45 GeV and 250 GeV. When applying SC, the distributions narrows slightly, and the high energy tail is reduced, following the general trend observed also for single neutral particles. As expected, the effect of SC is less pronounced in the case of jets.

	\begin{figure}[!htbp]	
  	\begin{center}	
    	\includegraphics[width=0.42\textwidth]{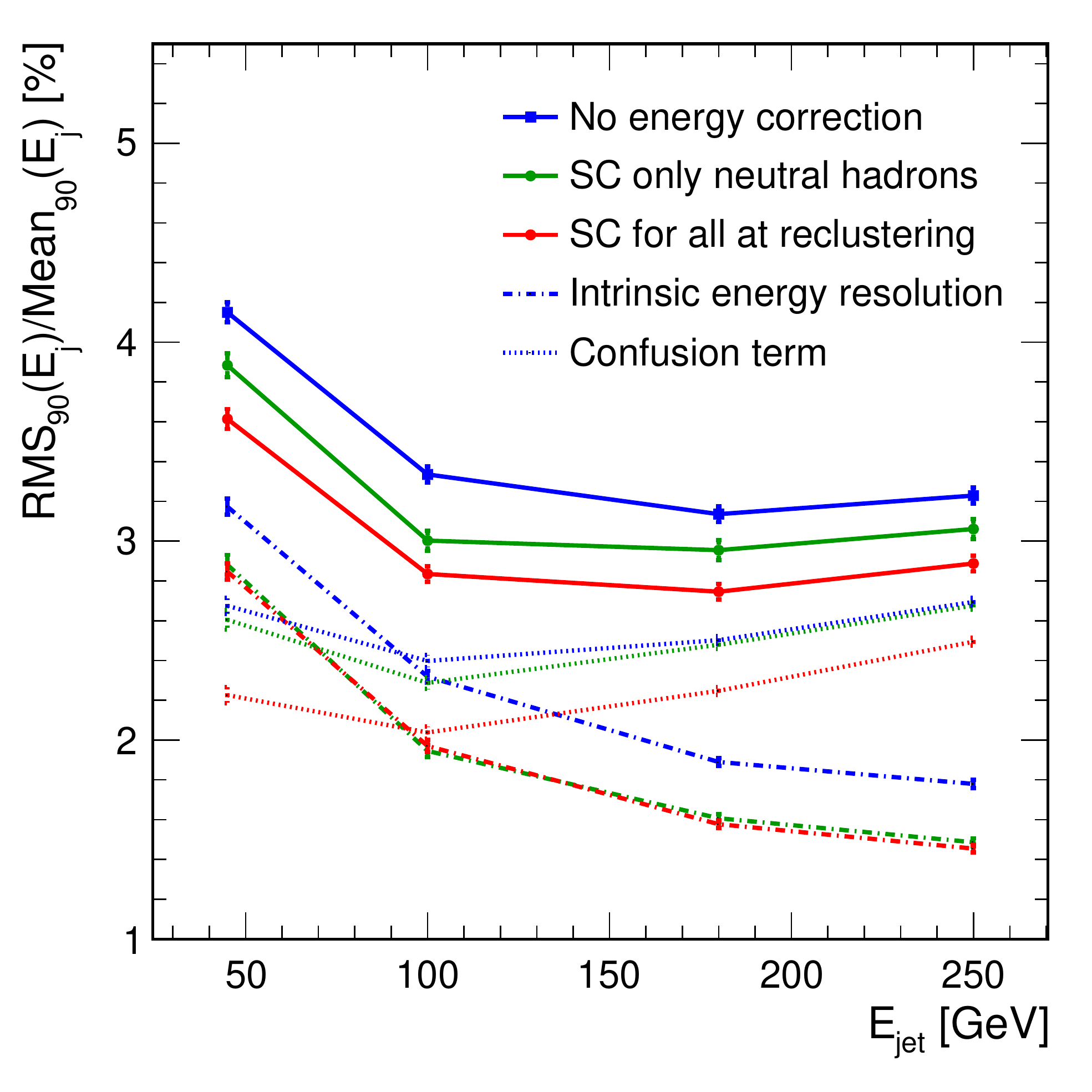}
  	\end{center}	
  	\caption{Jet energy resolution for the ILD detector with $3\times 3\ cm^{2}$ HCAL cell size without energy correction (blue), when applying SC for neutral clusters at PFO level (green) and when applying SC at the reclustering step for all clusters (red). The long-dashed lines present the intrinsic energy resolution, the dotted lines present the confusion term.}
  	\label{fig:JER_SCStages}
	\end{figure}

In Fig.~\ref{fig:JER_SCStages}, the jet energy resolution obtained for all three scenarios of the reconstruction, without energy correction, when applying SC for neutral clusters at the PFO level only and when applying SC also at the reclustering step for all clusters, is compared. The intrinsic energy resolution and the confusion term, derived from "cheated" reconstruction using perfect pattern recognition as discussed in Section \ref{subsection:intro_PFA} are also included in the figure. The application of SC only for neutral hadrons after the reclustering step only improves the intrinsic energy resolution without impact on the confusion term and thus mostly affects the lower part of the jet energy range. However, the application of SC in the reclustering step gives a substantial improvement of both contributions to the jet energy resolution, thus also affecting the high energy region and highlighting the two effects of a full integration of SC into the particle flow algorithm:
\begin{itemize}
	\item The improvement of the individual cluster energy measurement provides more powerful cluster-track compatibility constraints and therefore leads to confusion mitigation. This cannot be achieved with SC applied only at the PFO level.  
	\item The total energy distribution of a jet event is the sum of all single particle energies. The improved single particle resolution and a more gaussian energy distribution results in an intrinsically better jet energy resolution.  
\end{itemize}

\subsection{Alternative implementations of energy correction}
\label{subsection:discussions}

In the following, two alternative implementations of the energy correction in PandoraPFA are discussed. The first is the "standard" correction mechanism, which uses a truncation of the cell energy. The second uses the software compensation technique discussed above, making use of a different parametrisation of the hit energy weights, inspired by the semi-digital energy reconstruction developed for highly granular calorimeters with gas detector readout developed within the CALICE collaboration \cite{Buridon:2016ill}.

\subsubsection{Cell energy truncation}
\label{subsubsection:cell_energy_truncation}

As discussed in Section~\ref{subsection:implementation}, the earliest point where one can apply SC in particle flow is at the reclustering step. However, if the total cluster energy is not used to determine cell weights, an energy correction can already be applied at the hit level by simply reducing the response of the calorimeter to the hits that are likely to belong to electromagnetic sub-showers. For that, a simple cut on the cell energy is applied, limiting the response of individual cells. This technique is referred to as \textit{cell energy truncation}, and has been used as the default energy correction in PandoraPFA in the ILD detector concept.

	\begin{figure}[!htbp]	
  	\begin{center}	
    	\includegraphics[width=0.42\textwidth]{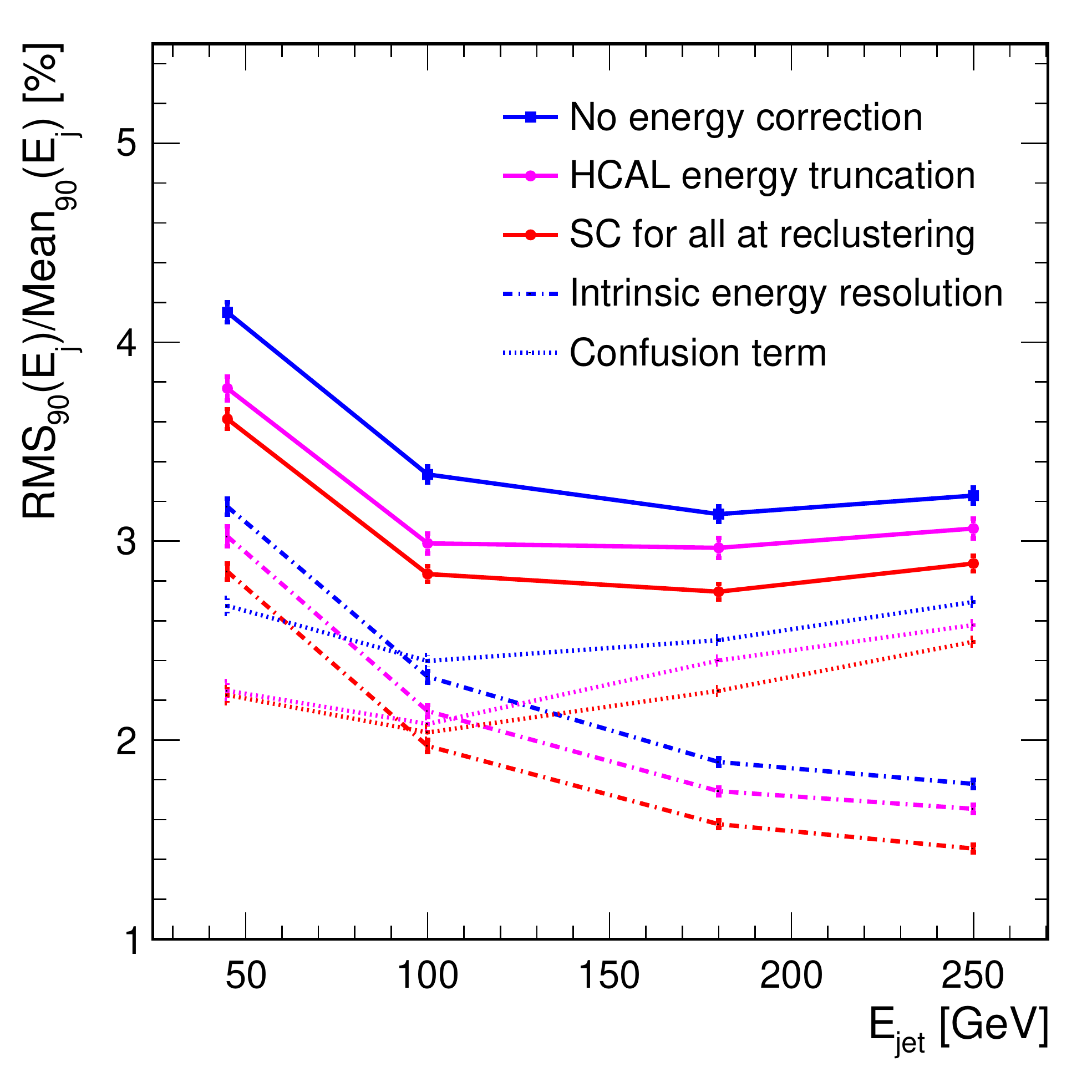}
  	\end{center}	
  	\caption{Jet energy resolution for ILD detector with standard HCAL cell size without energy correction (blue), when applying HCAL cell truncation combined with a downscaling of clusters with very high energy density (magenta) and when applying SC at reclustering step for all clusters (red). The long-dashed lines present the performance obtained with perfect pattern recognition, whereas the dotted lines present the confusion term, defined as the quadrature difference between the real jet energy resolution and the one obtained with perfect pattern recognition.}
  	\label{fig:JER_SCTruncation}
	\end{figure}

The advantage of the method over the more sophisticated software compensation lies in its simplicity. Nevertheless, the determination of the optimal cell energy truncation value is not trivial. The overall calibration also needs to be re-optimised once truncation is applied, as it reduces the total reconstructed energy. In PandoraPFA, the cell energy truncation method is applied in combination with a simple treatment of hadron showers with very high electromagnetic fraction (\textit{"hot hadrons"}): the energy of hadrons with a low number of hits and very high energy density is scaled down. The hot hadron treatment is in fact a simple form of compensation. With this combined method the jet energy resolution is improved significantly compared to the case without energy correction, as demonstrated in Fig.~\ref{fig:JER_SCTruncation}. However, the level of  improvement is smaller than for full SC at the reclustering step.

\subsubsection{Software compensation weights inspired by semi-digital energy reconstruction}
\label{subsubsection:sc_and_semidigital}

In a semi-digital hadron calorimeter, each cell provides only a 2-bit energy information corresponding to three hit energy thresholds, rather than the 12-bit "analogue" energy information available in the calorimeter considered here. To recover response linearity and energy resolution also at higher energies, the reconstructed energy is given by the sum of the hits above each of the three thresholds, each weighted with a parameter depending on the total number of hits in the shower. This dependence is parametrised by a second-order polynomial \cite{Buridon:2016ill}, and determined by a minimisation procedure identical to the one used here, described by Eq. \ref{eq:chi2_SC_weights}. 

In order to establish a general formalism covering both analogue and semi-digital reconstruction, as a first step the analogue software compensation has been implemented in a scheme in which the total energy dependence of the weights is parameterised in each hit energy density bin independently. This approach removes the enforced exponential dependence of the weights as a function of hit energy density. This monotonically falling weight towards higher energy densities is motivated by the higher density and higher detector response of electromagnetic sub-showers. Encoding this in the software compensation weights however does not necessarily result in the best possible performance. In this alternative scheme, the reconstructed energy is given by

\begin{equation}
\label{eq:SD_energy_computation}
\mathrm{E}_{\mathrm{SC}}  =  \sum _{\mathrm{bins} j} \ \sum_{\mathrm{hits}\, i\,\in\,\mathrm{bin}\, j} \alpha_{j} (\mathrm{E}) \mathrm{E}_{i},
\end{equation}
where the energy dependence of $\alpha$ is parameterised as a second order polynomial with 3 free 
parameters per bin to be optimised. 
This becomes semi-digital reconstruction, i.e.\ counting of hits in each of the bins, by replacing
$\alpha$ by $\alpha / \mathrm{E}_i$ in a second step, thus canceling the hit energy dependence, and expressing E as a function of 
the total number of hits.
The difference between software compensation and semi-digital reconstruction then reduces to a 
smaller number of bins and a forced effective  $1/\mathrm{E}_{\mathrm{hit}}$ dependence of the weight within each bin 
instead of weights being bin-wise constant. 
This implementation of the SC formalism thus prepares Pandora for the optimisation for 
calorimeters with semi-digital read-out. 

In order to verify the equivalence of the formalism for analogue software compensation, the binning in hit energy density was kept unchanged, and the weight parameters (here the parameters of the second-order polynomial for each bin) are determined by using Eq. \ref{eq:chi2_SC_weights}.

	\begin{figure}[!htbp]	
  	\begin{center}	
    	\includegraphics[width=0.42\textwidth]{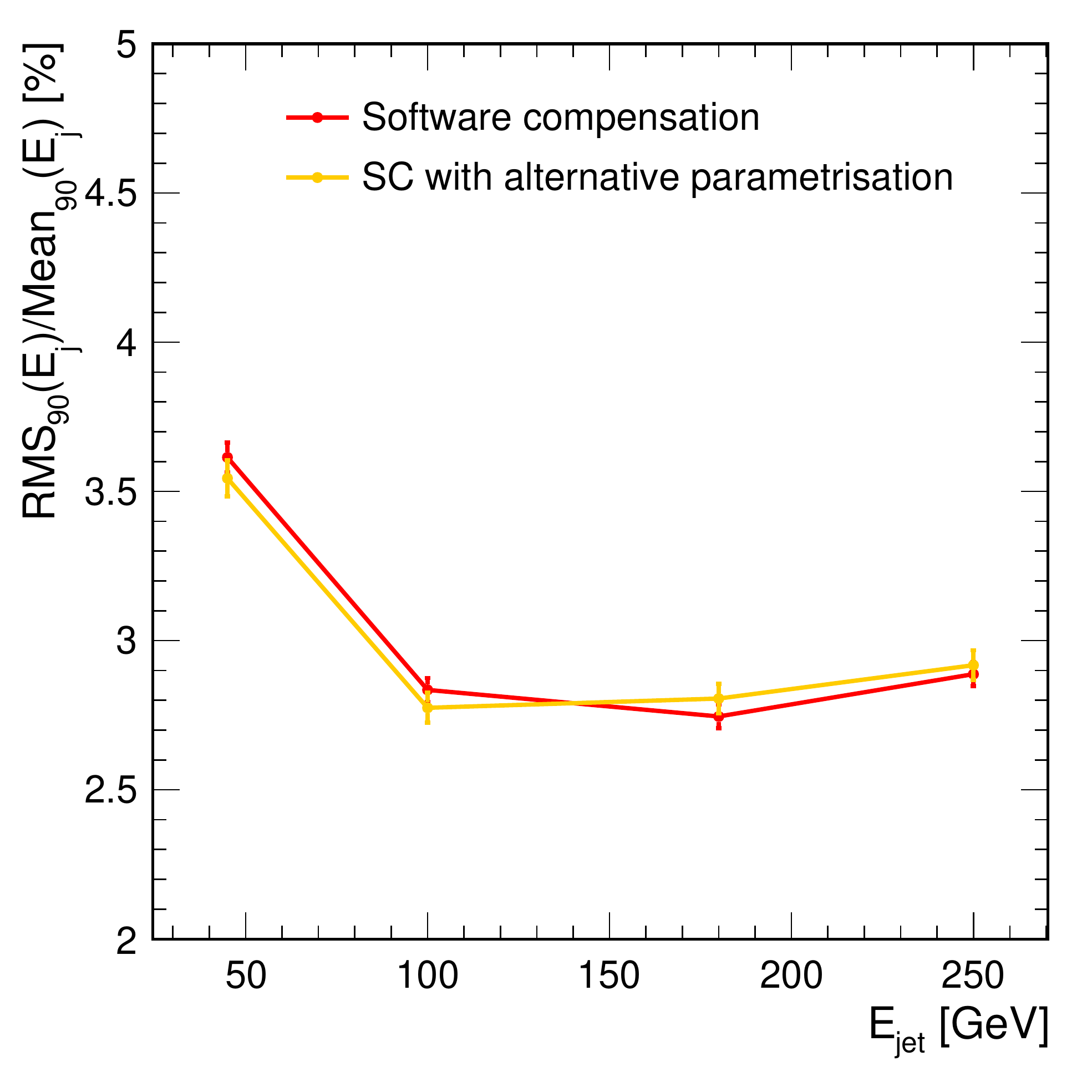}
  	\end{center}	
  	\caption{Jet energy resolution for SC applied at the reclustering stage for two different parametrisations of the cell energy weights, the exponential weight parametrisation (red) and the alternative parametrisation inspired by semi-digital energy reconstruction (orange).}
  	\label{fig:JER_AlternativePar}
	\end{figure}
	
Fig.~\ref{fig:JER_AlternativePar} shows the jet energy resolution obtained with SC fully applied in the reconstruction for both the standard and the semi-digital-inspired weight parametrisation. The performance of both cases is comparable for the range of energy considered, showing that the benefits of software compensation do not depend strongly on the details of the implementation of the hit weights.

\section{The influence of transverse granularity}
\label{section:AHCAL_cell_size_optimisation}

For a particle flow calorimeter, the level of granularity has an important influence on the performance, since it determines the possibilities for pattern recognition. Also the software compensation technique discussed here benefits from high granularity. Therefore the inclusion of software compensation methods in the particle flow reconstruction could in principle affect the choice of the optimal cell size. The present study is the first one to take this into account.  

The optimal granularity is an interplay of the physics performance and the complexity and cost of the detector. The latter has several components which are proportional to the detector's volume, the instrumented area and the number of channels. If the size and the number of layers of the detector are kept constant, the number of channels is the relevant quantity to be optimised. A smaller cell size may improve the jet energy resolution, but the corresponding larger number of cells results in an increased cost due to increased electronics requirements as well as a larger number of photon sensors in the case of scintillator-based calorimeters.  

	\begin{figure*}[!htbp]	
  	\begin{center}	
    	\includegraphics[width=0.84\textwidth]{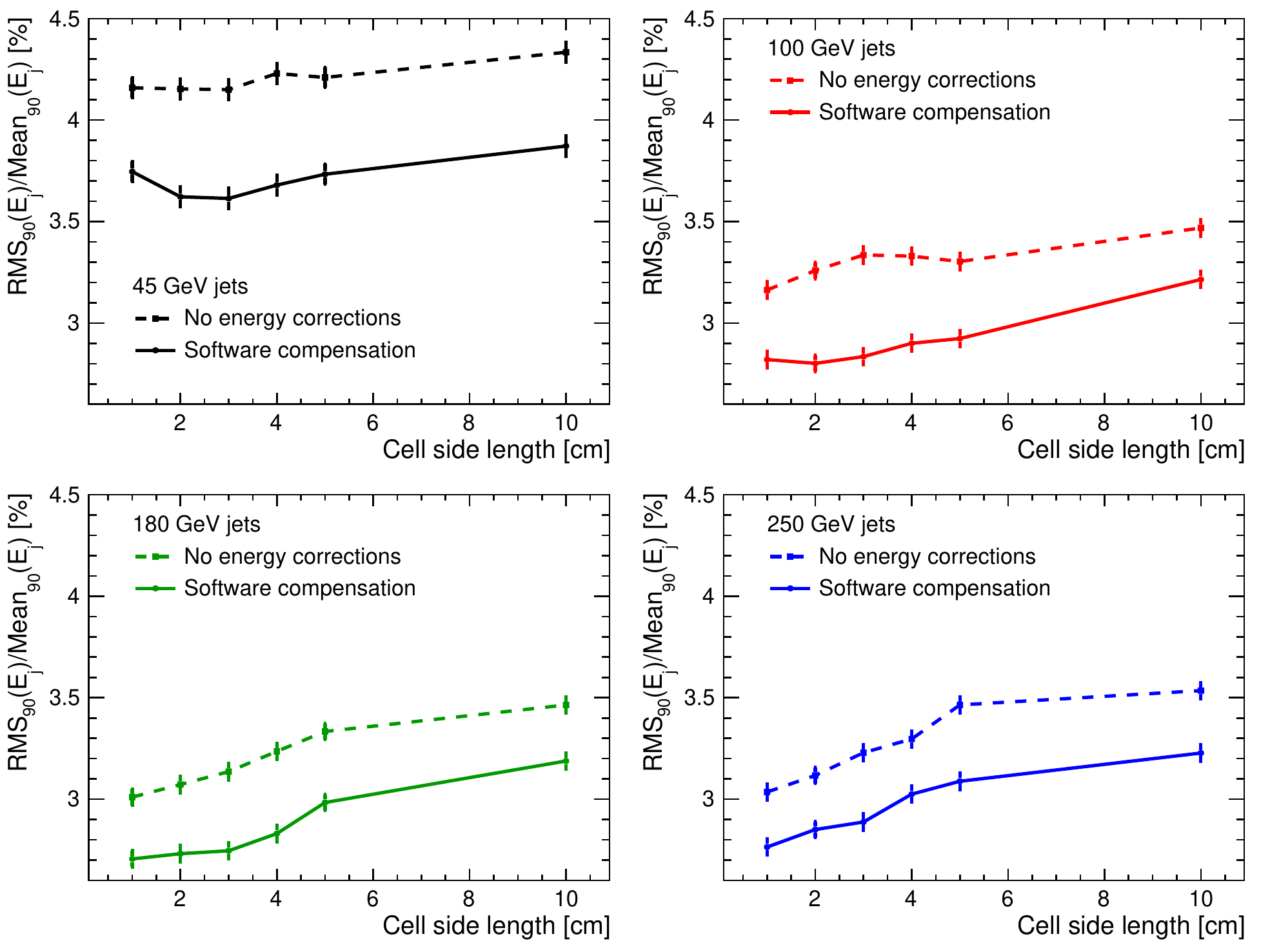}
  	\end{center}	
  	\caption{Jet energy resolution as a function of the HCAL cell size for different jet energies. The dashed lines show the jet energy resolution without energy correction. The solid lines show the jet energy resolution obtained when applying SC at the reclustering step.}
  	\label{fig:JER_vs_CellSize}
	\end{figure*}

In this study, the dependence of the jet energy resolution obtained with PandoraPFA on the cell size of the HCAL is studied for quadratic cells ranging from a size of $1\times 1$ $\mbox{cm}^{2}$ to \mbox{$10\times 10$ $\mbox{cm}^{2}$}, keeping all other parameters of the calorimeter constant. The study is performed with full application of software compensation and without any energy corrections as a comparison. When using software compensation, the weights to be used by the algorithm are determined separately for each cell size following the procedure described in section \ref{subsection:SC_weight_definition}. For consistency, the same  binning in energy density as well as the same noise rejection thresholds on the cell level are used for all detector configurations. Fig.~\ref{fig:JER_vs_CellSize} shows the resulting energy resolution as a function of cell size for the four different jet energies considered here. The application of software compensation leads to a significant improvement of the jet energy resolution for all cell sizes and all jet energies. The general trend of the evolution of the energy resolution with changing cell size is the same for reconstruction with and without the use of software compensation. The reduced improvement of the energy resolution with software compensation observed for the smallest cell sizes for lower jet energies is due to a non-optimal use of the lower energy density bins in these models. In combination with the noise rejection cut on cell level this results in a slight reduction of the performance of software compensation at lower jet energies, visible in particular for 45 GeV jets for the model with  $1\times 1$ $\mbox{cm}^{2}$ cells.

	\begin{figure}[!htbp]	
  	\begin{center}	
    	\includegraphics[width=0.42\textwidth]{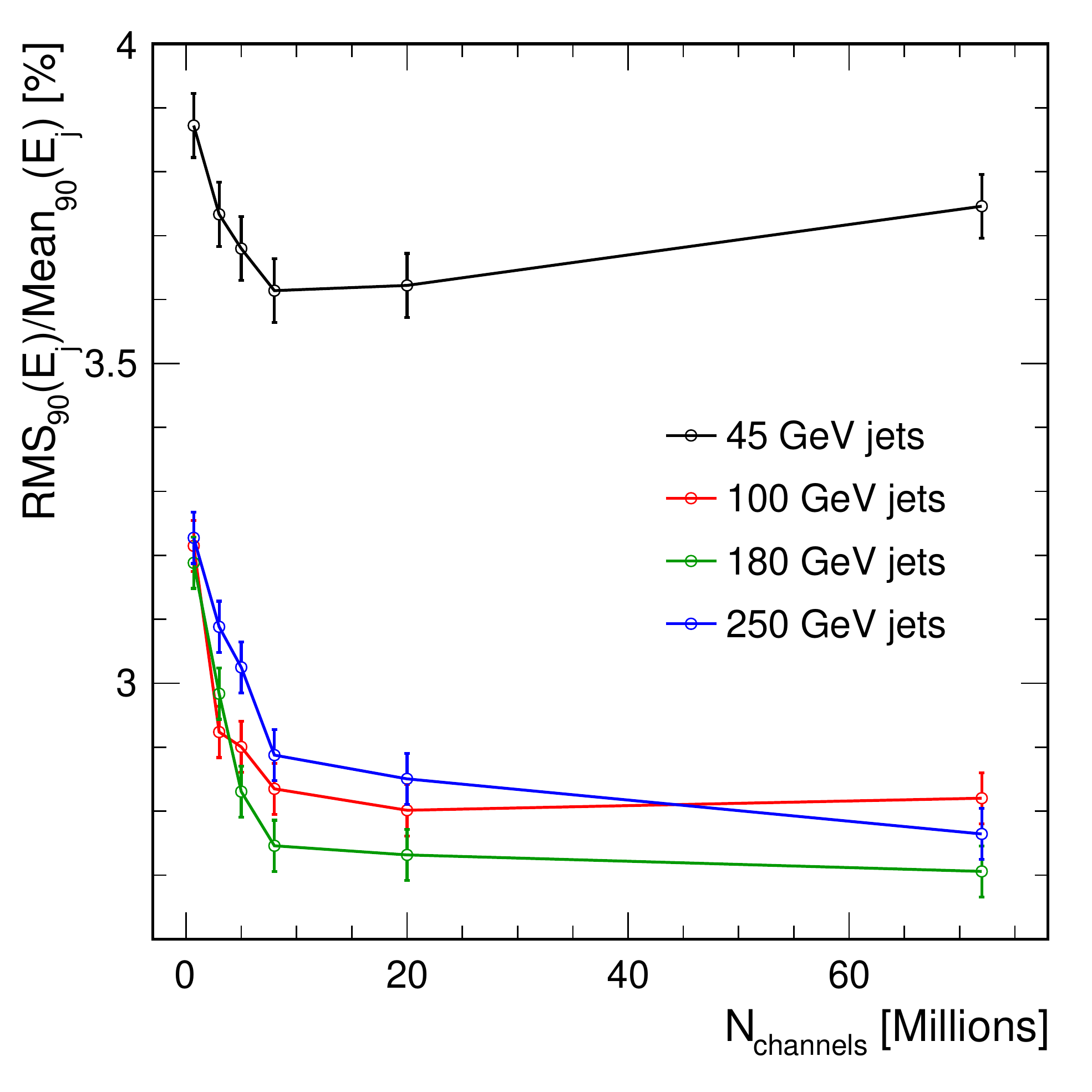}
  	\end{center}	
  	\caption{Jet energy resolution with software compensation as a function of number of channels in the HCAL for different jet energies.}
  	\label{fig:JER_vs_Nchannels}
	\end{figure}
	
As mentioned above, the optimisation of the cell size of the detector has to find a balance between physics performance and and the number of readout channels, which is expected to influence the detector cost. To inform this choice, the results shown in Fig.~\ref{fig:JER_vs_CellSize} are shown as a function of the number of HCAL cells in the full ILD detector in Fig.~\ref{fig:JER_vs_Nchannels}. Note that the HCAL baseline design for ILD consists of about 8 millions cells with a size of \mbox{$3\times 3$ $\mbox{cm}^{2}$}. The figure illustrates  that the HCAL baseline design represents a well-motivated choice, being located at the knee of the distribution. A larger cell size (corresponding a smaller number of cells) results in a noticeable degradation of the jet energy resolution, while a smaller cell size does not provide substantial improvement, albeit very significantly increasing the number of cells.

Beyond the change of the size of all cells in the calorimeter, an overall reduction of the channel count can also be achieved by using larger cells only in the rear part of the detector, where pattern recognition requirements may be less strict due to reduced occupancy. To investigate this scenario, a study has been performed that uses two different cell sizes, $3\times 3$ $\mbox{cm}^{2}$ in the front part of the detector and $6\times 6$ $\mbox{cm}^{2}$ in the rear part. Since the consistent implementation of software compensation in a detector with varying granularity goes beyond the scheme developed here, this study is performed with the standard reconstruction without software compensation or any other form of energy correction. 

	\begin{figure}[!htbp]	
  	\begin{center}	
    	\includegraphics[width=0.42\textwidth]{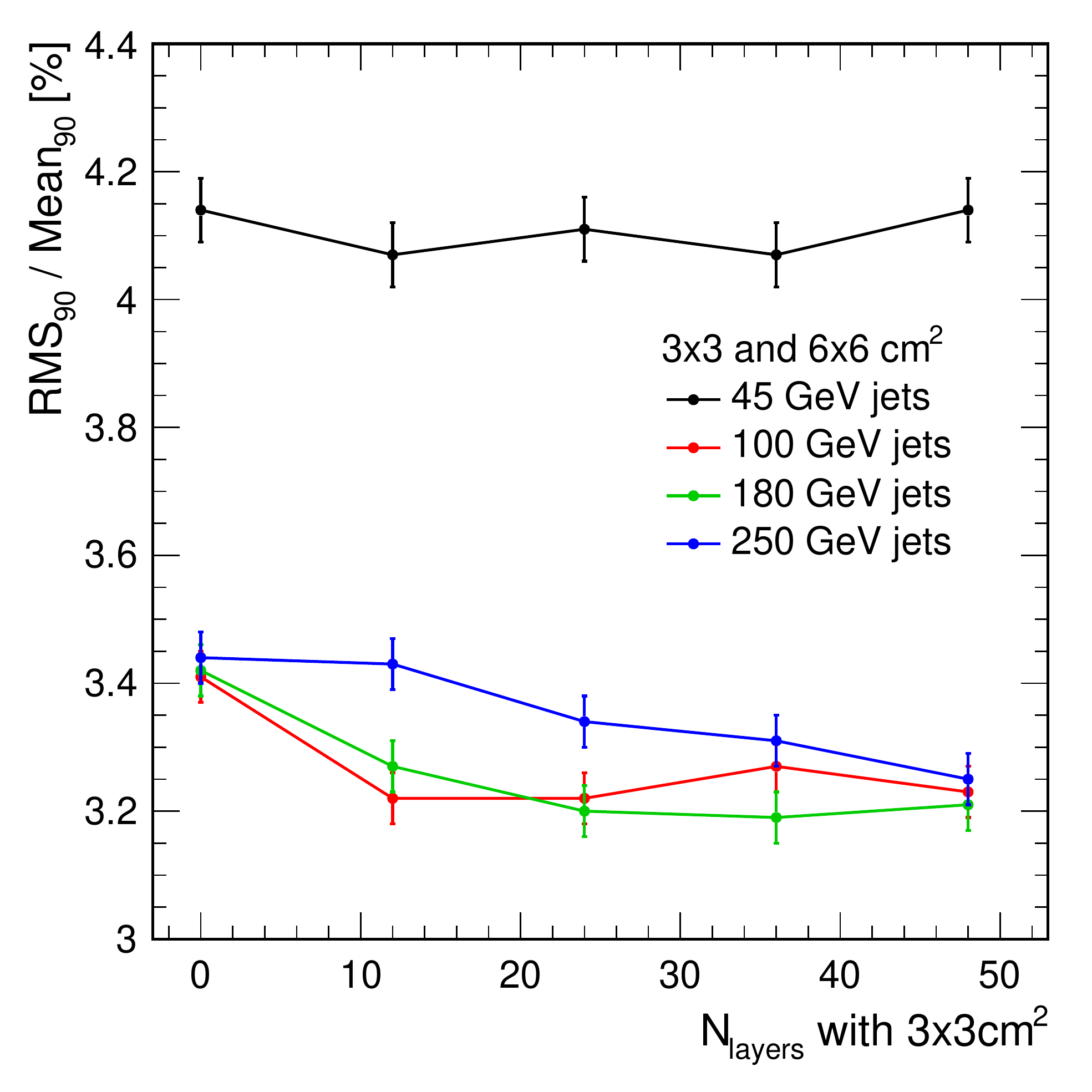}
  	\end{center}	
  	\caption{Jet energy resolution with standard reconstruction (no software compensation, no energy truncation) as a function of the number of high granularity layers in the front part of the HCAL for different jet energies. Here, 0 corresponds to $6\times 6$ $\mbox{cm}^{2}$  throughout the full calorimeter, while 48 represents the scenario with $3\times 3$ $\mbox{cm}^{2}$ granularity in the full system studied in the main part of this paper.}
  	\label{fig:JER_vs_TransitionLayer}
	\end{figure}

Figure \ref{fig:JER_vs_TransitionLayer} shows the jet energy resolution as a function of the number of highly granular layers in the front part of the hadron calorimeter for the standard reconstruction without software compensation for different jet energies. The impact of the depth at which the transition from a fine to a coarse HCAL occurs depends on the jet energy. While there is only a small impact for low energies, higher energies prefer a deeper location of the transition, with 250 GeV jets still improving when also implementing the higher granularity in the last 10 layers.

\section{Conclusions}
\label{section:conclusions}

The software compensation technique developed for the CALICE analogue hadron calorimeter, making use of local energy density information within hadronic showers to substantially improve the energy resolution for single hadrons, has been integrated into particle flow event reconstruction based on the PandoraPFA algorithm. The performance of this algorithm is studied on simulated single particle and di-jet events in the ILD detector concept for the International Linear Collider. 

The results demonstrate that the application of software compensation in particle flow reconstruction in the HCAL substantially improves the detector performance, both due to the improvement of the energy measurement of neutral hadrons and an improved track-cluster matching due to more accurate calorimeter energy information. For single neutral hadrons, the improvement of the  energy resolution is compatible to the result previously obtained within the CALICE collaboration using test beam data. For the jet energy resolution, in addition to the improvement of the intrinsic particle energy resolution, the application of software compensation also results in a reduction of  the confusion term. The ILD goal of a jet energy resolution of better than 4\% over the full relevant energy range is achieved, with a jet energy resolution of better than 3.5\% obtained for energies higher than 45 GeV. The full software compensation outperforms less sophisticated energy correction techniques such as a truncation of very high energy depositions in single calorimeter cells.  An alternative implementation of the SC compensation formalism, shown to provide 
identical performance for the analogue HCAL, prepares Pandora for the application to calorimeters 
with semi-digital read-out. 

PandoraPFA with integrated software compensation has also been used to revisit the optimisation of the cell size of the scintillator / SiPM - based analogue HCAL of the ILD detector concepts. The results confirm that a size of $3 \times 3$ $\mbox{cm}^{2}$ is an optimal choice for the scintillator tiles considering performance and overall detector channel count. A channel count reduction can also be achieved by making only the rear part of the hadron calorimeter less granular. The point at which such a transition does not negatively impact the resolution depends on the jet energy, with the highest energies studied here favouring high granularity throughout the detector. 

The results presented in this paper demonstrate that the application of local software compensation in the analogue hadron calorimeter has an important impact on particle flow reconstruction and improves energy resolution for single particles and jets. Following the present study, software compensation is used as the standard energy correction algorithm in PandoraPFA version v02-09-00 and later versions for physics studies of ILD.

\section*{Acknowledgments}
We would like to thank our CALICE and ILD colleagues for their excellent collaboration and fruitful discussions. This study would not have been possible without making extensive use of the ILD simulation and reconstruction software framework. This project has received funding from the European Union's Horizon 2020 Research and Innovation programme under Grant Agreement no. 654168 and was supported by the UK Science \& Technology Facilities Council and by the DFG cluster of excellence `Origin and Structure of the Universe'.

\bibliography{References}

\end{document}